# Waves and instability in a one-dimensional microfluidic array

Bin Liu, J. Goree, and Yan Feng
*Department of Physics and Astronomy, The University of Iowa, Iowa City, Iowa 52242*
(Dated: June 21, 2012)

Motion in a one-dimensional (1D) microfluidic array is simulated. Water droplets, dragged by flowing oil, are arranged in a single row, and due to their hydrodynamic interactions spacing between these droplets oscillates with a wave-like motion that is longitudinal or transverse. The simulation yields wave spectra that agree well with experiment. The wave-like motion has an instability which is confirmed to arise from nonlinearities in the interaction potential. The instability's growth is spatially localized. By selecting an appropriate correlation function, the interaction between the longitudinal and transverse waves is described.

PACS numbers: 47.55.D-, 47.60.-i, 47.20.Ky, 63.22.-m

## I. INTRODUCTION

Flows at a micron scale, called microfluidic flows, are of interest in fields such as molecular analysis, molecular biology and microelectronics [1]. Some microfluidic flows include a dispersed phase, such as droplets, bubbles, biological cells or colloidal particles. In microfluidics, there is a flow in a channel that has at least one dimension that is hundreds of microns or smaller. In such a channel, a dispersed phase such as water droplets can be made to align in an array consisting of a single row, as sketched in Fig. 1; structures like these have been termed a 1-D array [2, 3], a 1-D stream [4], or a 1-D crystal [5–7]. Such 1-D arrays have applications in protein crystallization [2, 3] and flow cytometry [9], and they can exhibit physical phenomena such as waves [4–7] and instabilities [5–7, 10].

In this paper, we report a numerical simulation that mimics the experiment of Beatus et al. [5]. In that experiment, a 1-D array of water droplets was dragged by a flow of oil in a microfluidic channel, as sketched in Fig. 1. Due to the friction they experience on the floor and ceiling of the channel, the water droplets move more slowly than the oil. As they were injected into the oil flow, the water droplets at first self-organized in an equilibrium along the centerline of the channel, and they were positioned in a single row with a highly regular spacing, and the authors described this regularly-spaced array of droplets as a 1-D crystal. One discovery in the experiment of Beatus et al. was that the droplets fluctuate about their equilibrium positions with a wave-like motion and these oscillations are surprisingly not overdamped by viscous dissipation. This wave-like motion, termed phonon by the experimenters, included displacements in two directions with respect to the flow, parallel and perpendicular, corresponding to longitudinal and transverse waves, respectively. Another finding in the experiment was an instability that developed farther downstream. In this instability, a local fluctuation in the 1-D array increased with time. As this instability grew, the displacements of droplets developed large amplitude fluctuation, which the experimenters attributed to nonlinear effects due to two factors: large amplitude motion and an interaction between longitudinal and transverse waves. Eventually the 1-D structure was destroyed by an extreme transverse displacement of droplets that allowed them to move past another in the longitudinal direction. The energy source for the instability was the flow of the oil through the channel [5].

Both the wave motion and instability in the dispersed phase were described by Beatus et al. [5–7] as the result of hydrodynamic interaction. One droplet disturbs the surrounding flow, which in turn disturbs another droplet via a drag force on the droplet due to the faster oil. Manifestations of this kind of hydrodynamic interaction have been reported for other microfluidic flows; these manifestations include self-assembly patterns [11–13], shock waves [4, 7], and oscillatory waves. The latter were discovered experimentally by Beatus et al. [5, 7] and studied in simulations where the flow field scattered by droplets or particles is modeled as a superposition of the far-fields of dipoles [5, 7] or is calculated by a Stokesian-dynamics method [8].

Our simulation treats individual droplets as particles that move according to a well-defined equation of motion. Instead of computing the entire flow field for the oil, we model the hydrodynamic interactions due to the flow field in the droplets' equation of motion. The model we use for the hydrodynamic interactions was developed theoretically by Beatus et al. [5, 7] who treated the oil flow as a potential flow.

The simulation results we report here show agreement with the wave spectrum, dispersion relation, and instability in the experiment of Beatus et al. [5]. We use the simulation to verify that the instability is due to nonlinearity, as proposed by Beatus et al. [5]. We use the simulation results to describe the interaction of the longitudinal and transverse waves by selecting an appropriate correlation function based on the time series of the longitudinal and transverse microscopic currents.

The physics that will be studied here, in a 1-D microfluidic array, is also relevant to other physical systems that have a single row of particles or atoms. These systems include a line of colloidal microspheres confined by holographic optical traps [14], a line of charged micron-sized particles levitated in dusty plasmas [15, 16], and a line of helium atoms adsorbed on bundles of single-walled carbon nanotubes [17]. In these examples, the particles require some mechanism to confine them so that they form a 1-D array, and the collection of particles can sustain wave-like motion, which is due to a combination of the confinement and the interaction between the particles.

## II. SIMULATION

We simulate a 1-D array of water droplets with the same configuration and parameters as in the experiment of Beatus et al. [5]. As sketched in Fig. 1, the channel is aligned in the $x$ direction and it has a rectangular cross-section of width 250 $\mu$m and a height 10 $\mu$m that is the same as the droplet's radius $R = 10$ $\mu$m. The simulation begins after water droplets have been injected, so that they are initially spaced by $a = 27$ $\mu$m. In calculating the forces acting on the droplets, we assume an oil flow velocity of $u_{oil}^\infty = 1730$ $\mu$m/s, while the droplets move more slowly, at $u_d = 380$ $\mu$m/s.

We numerically integrate the equations of motion of all the simulated droplets. A droplet is treated individually as a point object that responds to hydrodynamic forces, which are calculated to take into account of the finite radius $R$ of the flattened droplet. The equation of motion for the $n^{th}$ droplet is [5]

$$\frac{d\mathbf{r}_n}{dt} = \frac{u_d}{u_{oil}^\infty} \nabla \phi(\mathbf{r}_n). \quad (1)$$

This equation of motion expresses a balance of two forces: the friction on the channel walls and the drag in the oil flow. The equation of motion in Eq. (1) includes only hydrodynamic effects; other effects such as Brownian motion due to the discreteness of molecules in the oil or water are neglected. The inertial term in Newton's second law is also neglected, so that unlike some equations of motion, Eq. (1) is a first-order differential equation. For each droplet, we integrate the equation of motion, so that the position $\mathbf{r}_n$ and velocity $d\mathbf{r}_n/dt$ of all the droplets are advanced simultaneously in a sequence of time steps.

As in Ref. [5], we model the potential $\phi$ as

$$\phi(\mathbf{r}) = u_{oil}^\infty x + \sum_m \phi_d(\mathbf{r} - \mathbf{r}_m), \quad (2)$$

where the first term on the right is a potential due to oil flowing uniformly at $u_{oil}^\infty$ and the second term is a superposition of the potentials due to all other droplets $m$. An underlying assumption in Eq. (2) is a pairwise interaction of droplets, meaning that the droplet phase is sufficiently dilute that three-body interactions can be neglected. For such a dilute phase, it is appropriate to model the far-field interaction using only a dipole potential. As in [5], for the interaction potential $\phi_d$ we use

$$\phi_d(r) = R^2(u_{oil}^\infty - u_d)\frac{x}{r^2}, \quad (3)$$

which is the potential due to a dipole aligned with the flow in the $\hat{x}$ direction. Here, $r = \sqrt{x^2 + y^2}$ is the distance from the droplet located at $x = 0$, $y = 0$. Equation (3) assumes that the droplet has been flattened by the channel to have the shape of a thin disc, and that the surrounding flow has a Poiseuille-like parabolic profile along the $z$ axis and can be described in the $xy$ plane as a potential flow, as in a Hele-Shaw cell. Equation (3) also assumes that the droplets are unconfined in the $xy$ plane, i.e., we assume that the channel width is so great that its effects can be neglected. Such finite width effects could be accounted for by using a different potential [6, 7].

The force, which varies as $-\nabla\phi$ in the right-hand-side of Eq. (1), is nonlinear with respect to displacements of a droplet, when using Eqs. (2) and (3) to describe $\phi$. Most of our results are for this nonlinear case, although in Sec. III C 3 we will perform a test with a linearized form of $-\nabla\phi$. The latter is obtained by performing a Taylor expansion of Eq. (3) for small displacements $\delta x$ and $\delta y$ from the equilibrium position $x_0$, where $x = x_0 + \delta x$ and $y = \delta y$. This expansion of Eq. (3) is

$$\phi_d(x,y) \approx \frac{R^2(u_{oil}^\infty - u_d)}{x_0}[1 \ - \ \frac{\delta x}{x_0} + \left(\frac{\delta x}{x_0}\right)^2 \\ - \left(\frac{\delta y}{x_0}\right)^2 + \text{h.o.t.}]. \quad (4)$$

To turn off nonlinear effects in $-\nabla\phi$, we can retain only the first four terms in the brackets and neglect the higher-order terms (h.o.t.); we will do this only in Sec. III C 3. For all other results, we will use the fully nonlinear force computed using Eq. (3) instead.

Our simulation includes $N = 256$ droplets, and we use a periodic boundary condition in order to mimic an infinite 1-D array. The calculation is performed in the inertial frame of the droplets. To imitate the confinement provided by the floor and ceiling of the channel, the droplet is constrained to have no movement in the vertical direction, so that the equation of motion is integrated in only two coordinates, $x$ and $y$. The simulation runs for a duration of 10 s. To advance a droplet's position, we use a fourth-order Runge-Kutta integrator with a fixed time step [19]. The time step was selected as $5.6 \times 10^{-5}$ s by performing a test requiring that discrepancies in the kinetic energy of droplets are never significant over the simulation's duration.

Because the droplets in the 1-D array are allowed to move in two directions, perpendicular and parallel to the flow, the 1-D array can sustain two modes, longitudinal and transverse, respectively. Due to the finite number of droplets, small amplitude motion can be decomposed into 256 discrete sinusoidal modes for the longitudinal motion, and the same for the transverse motion. Thus, the allowed values of $ka$ in our simulation are $\pm\pi/128$, $\pm 2\pi/128$, ..., $\pm\pi$, where $a = 27\mu$m is the equilibrium spacing (lattice constant), and $k = 2\pi/\lambda$ is the wave number for a mode of wavelength $\lambda$.

Our initial conditions at $t = 0$ are chosen to mimic small random displacements of the droplets from their equilibrium positions. We start a droplet $m$ at a position $x_m = (m - N/2 + \delta_x)a$ and $y_m = \delta_y a$. Here, $\delta_x$ and $\delta_y$ are random numbers, with a mean of zero and a standard deviation of 0.01. The random numbers, which were different for each droplet, were chosen so that there would initially be equal spectral power in each of the 256 longitudinal and 256 transverse modes. (This was done by initially giving each mode the same amplitude, but with a random phase.)

As an approximation that is typical for molecular dynamics simulations, we cut off the potential at large radius, so that $\phi_d$ is given by Eq. (3) for $r \leq r_{\text{cut}}$, or it is zero for $r > r_{\text{cut}}$ [18]. To select our cutoff radius $r_{\text{cut}} = 64a$, we performed a test requiring that the kinetic energy never develop a significant discrepancy.

After running the simulation and obtaining time series of the positions and velocities of all droplets, we use time series data for the longitudinal position $x_m$ and the velocity $(v_{x,m}, v_{y,m})$ for each droplet to calculate

$$j_L(k,t) = \sum_{m=1}^{N} v_{x,m}(t) e^{-ikx_m(t)} \quad (5)$$

and

$$j_T(k,t) = \sum_{m=1}^{N} v_{y,m}(t) e^{-ikx_m(t)}. \quad (6)$$

Equations (5) and (6) describe the fluctuating velocities of droplets in the longitudinal ($\hat{x}$) and transverse ($\hat{y}$) directions, respectively, for a specified value of the wave number $k$. In the literature for the physics of liquids, $j_L$ and $j_T$ are sometimes called "microscopic currents" [20], which can characterize the oscillatory wave-like motion of molecules in a liquid. Here, instead of molecules, we track the motion of droplets. We compute these currents for all 256 longitudinal and 256 transverse modes that are allowed for our simulation.

### III. RESULTS

#### A. Droplet positions and velocities

In Fig. 2 we present raw simulation data consisting of positions recorded at three times. At $t = 0$ the simulation's initial conditions are shown, with droplets located almost at equal spacing $a$ along the centerline of the channel, with small random displacements in both $x$ and $y$, as described in Sec. II. At $t = 5$ s, the displacements of the droplets have become large enough to be easily detected by our diagnostics, although they are still too small to be identified in a visual inspection of Fig. 2. At the end of the simulation, $t = 10$ s, the displacements of the droplets have grown more, as large as $a$, due to the instability. At $x/a \approx -94$ in Fig. 2(c), we observe a fluctuation resembling the 1+3 arrangement of droplets observed in the experiment [5, 7] and the simulation of Beatus et al. [7]. After 10 s, the droplets move past one another, i.e., change their sequence, and at that point in the development of the instability, the 1-D lattice structure is essentially broken. The three panels shown in Fig. 2 are representative of the entire simulation movie, which we provide in the Supplemental Material [21].

We will use the raw data for droplet positions and velocities, and the currents computed from them using Eqs. (5) and (6), as the inputs to several diagnostics to characterize the waves and instability, as described next.

#### B. Waves

##### 1. Wave spectra

In Fig. 3 we present the wave spectra, which characterize wave-like motion in the 1-D array of water droplets. These spectra were prepared by computing fast-Fourier transforms (FFT) of the currents $j_L(k,t)$ and $j_T(k,t)$ with respect to time for each allowed $k$, and then plotting the spectral power (the square of the modulus of the FFT) as a function of $k$ and frequency $\omega$. This spectral power is physically related to the kinetic energy of the droplets.

We find that our wave spectra in Fig. 3 agree with the experimental spectra [5]. The most obvious difference is that our spectra do not include an artifact of unwanted fabrication defects in the channel that deflected the droplet motion in the experiment.

*a. Dispersion relation.* For both the longitudinal wave in Fig. 3(a) and the transverse wave in Fig. 3(b), the spectral power is significant only along a curve in the $\omega$ and $k$ parameter space. This variation of the resonance frequency with wave number indicates the wave dispersion relation.

For comparison, we also show in Fig. 3 the theoretical dispersion relation derived in [5] from a linearized equation of motion for small-amplitude motion with a dipole interaction

$$\begin{aligned} \omega_L(k) &= -\frac{6C_s}{\pi^2 a} \sum_m \frac{\sin(mka)}{m^3} \\ \omega_T(k) &= \frac{6C_s}{\pi^2 a} \sum_m \frac{\sin(mka)}{m^3}. \end{aligned} \quad (7)$$

We calculated the dispersion relation curves for this linear theory using the experimental values of the droplet spacing $a$ and sound velocity $C_s \simeq (2\pi^2 R^2 u_d/3a^2 u_{oil}^\infty)(u_{oil}^\infty - u_d)$, with no free parameters. We find that the dispersion relation for the linear theory accurately matches our simulation results.

The two waves propagate in opposite directions. In the inertial frame of the droplets, the transverse wave propagates in the $+x$ (i.e., in the same direction as the oil flow), while the longitudinal wave propagates in the $-x$ direction (opposite to the oil flow). These directions of propagation can be identified in Fig. 3 by noting that for small $|ka|$, the phase velocities $v_\phi = \omega/k$ of transverse and longitudinal waves are positive and negative, respectively. These wave directions are the same as was observed in the experiment [5].

We find that the power in our spectra is not equally partitioned among the possible values of wave number $k$. The power becomes concentrated at the wave number where the frequency is maximum, $ka \approx \pm\pi/2$. We will comment upon this wave number later.

*b. Feature not predicted by linear theory.* Although the dispersion relations observed in our simulation agree with the linear theory, we also observe a weaker feature that is not predicted by the linear theory of Eq. (7). This feature is seen near $ka \approx \pm 0.8$ and $\omega \approx \pm 10$ s$^{-1}$, where it is marked #3 in Fig. 3(b). Because these modes are not expected from a linear theory, there must be some nonlinear effects involved.

Beatus *et al.* [5] suggested that nonlinearity arose in their experiment due to two factors: large amplitude motion and an interaction of the longitudinal and transverse waves. Their suggestion motivates us to consider what type of wave-wave interaction it could be.

One possible type of nonlinear interaction is the generation of a third wave due by nonlinear mixing of two other waves obeying the criteria

$$\omega_1 + \omega_2 = \omega_3, \tag{8}$$

and

$$k_1 + k_2 = k_3. \tag{9}$$

Such a process is called "parametric decay" in plasma physics [22] and "three-wave" mixing in optics [23], and we will use the latter terminology. An argument in favor of this three-wave mixing hypothesis is that the $\omega$ and $k$ of the longitudinal waves marked #1 in Fig. 3(a) and the transverse waves marked #2 in Fig. 3(b), add up to match those of the feature marked #3 in Fig. 3(b), as in Eqs. (8) and (9). However, a counterargument is that feature #3 does not lie on an allowed dispersion relation for linear waves, unlike the best known cases of three-wave mixing in plasma physics [22].

### 2. Wave correlation

Having found in Fig. 3 a possible indication of nonlinear coupling of the longitudinal and transverse waves in the spectra, we now seek another diagnostic to indicate any interaction or synchronization of these two waves. Using the longitudinal and transverse currents, $j_L(k_L, t)$ and $j_T(k_T, t)$, respectively, we calculate a correlation function

$$C_{LT}(k_L, k_T, \tau) = \frac{\langle j_L(k_L, t) j_T(k_T, t+\tau)\rangle}{\sqrt{\left\langle |j_L(k_L,t)|^2 \right\rangle \left\langle |j_T(k_T,t)|^2 \right\rangle}} \tag{10}$$

In Eq. (10), $\tau$ is a time delay, and $\langle \cdots \rangle$ is an average over time $t$. The time interval for the averaging was selected to be as large as possible, given the 10 s duration of our simulation. The correlation function $C_{LT}$ is a complex number; we will report its modulus

$$|C| \equiv |C_{LT}(k_L, k_T, 0)|, \tag{11}$$

as a measure of the strength of the correlation between the two waves. For steady conditions, a value $|C| = 1$ would indicate perfect correlation (although we should note that the conditions here are not steady, due to the growth of the instability). Since we are investigating whether a longitudinal wave and a transverse wave are correlated, we present $|C|$ as a function of the longitudinal wave number $k_L$ and transverse wave number $k_T$, Fig. 4.

The results in Fig. 4 indicate a significant correlation of the longitudinal and transverse waves. This is evident from the dark regions in Fig. 4, which indicate a high correlation $|C|$.

To help interpret the correlation data in Fig. 4, we present Fig. 5 to indicate the conditions of highest correlation. We find that high correlations occur for waves that obey a matching condition for wave number, which is either

$$k_L = -k_T \tag{12}$$

or

$$k_L a = k_T a \pm (\pi - \epsilon), \tag{13}$$

where $\epsilon \ll 1$.

We also find that the waves that exhibit high correlation have a matching condition for frequency

$$\omega_L = \omega_T. \tag{14}$$

To demonstrate this matching condition, we combine the dependence of our correlation on wave number and the theoretical dispersion relations, Eq. (7), which are graphed on the edges of the main panel of Fig. 5. In examining Fig. 5, one can start with the open circle on the longitudinal dispersion relation, and then follow dashed line $H1$ to the left, where it is seen to have high correlations with two wave numbers for transverse waves, as indicated by vertical lines $V1$ and $V2$. All three of these modes (indicated by open and solid circles) have frequencies that are identical. This observation leads us to the frequency matching condition, Eq. (14).

While our correlation function indicates a significant correlation, it does not explain the mechanism for this correlation. This is, of course, a common limitation of using correlation functions.

### 3. Wave fronts

To visualize the spatiotemporal characteristics of the wave fronts, we present in Fig. 6 a space-time diagram,

which is a plot of droplet velocity as a function of position $x$ and time $t$. We prepared this space-time diagram by combining the time series for position $x(t)$ and one component of the velocity, for example $v_x(t)$, to generate contours of constant $v_x$ in the parameter space $x$ vs. $t$. A darker shade in Fig. 6 indicates a higher droplet speed. These space-time diagrams are useful for characterizing the spatiotemporal development of the droplet motion over the course of the entire simulation.

Features that can be seen in the space-time diagram include wave fronts for the longitudinal and transverse waves. These wave fronts appear as sloped stripes. Because the longitudinal waves propagate in the $-x$ direction, while the transverse waves propagate in the $+x$ direction, these two kinds of wave fronts are sloped oppositely in Fig. 6.

The observed phase velocity $v_\phi$ of the waves is found by measuring the slope of the wave fronts in the space-time diagram. We find that $v_\phi = -100$ $\mu$m/s for the longitudinal and $v_\phi = 100$ $\mu$m/s for the transverse waves.

### C. Instability

We now present our results for the instability. We will find that the instability grows nonexponentially, with a distinctive spatial localization. We will also confirm that the instability requires: (1) nonlinearity in the forces and (2) a coupling between longitudinal and transverse motions.

#### 1. Temporal growth

A feature that can be seen in the space-time diagram of Fig. 6 is a growth trend for the amplitude of the waves. As time progresses, the velocity amplitudes of both the longitudinal and transverse waves grow, as indicated by the shading in Fig. 6 appearing darker at larger times.

To reveal the scaling of the growth with time, we can also examine a time series of the kinetic energy averaged over the entire length of the 1-D array, which we present in Fig. 7. We find that the kinetic energy grows with time, but unlike some hydrodynamic instabilities such as the Rayleigh-Taylor instability, this one has a growth that is not exponential with time.

#### 2. Spatial localization

In addition to its temporal dependence, we can also characterize the spatial dependence of the instability's growth. This can be examined in the space-time diagram, Fig. 6. We find that the instability does not grow uniformly along the entire length of the 1-D array, but instead it develops in the form of isolated fluctuations. An example of this spatial concentration is seen prominently at $t > 8$ s, as marked $*$ in Fig. 6(b). This large-amplitude disturbance is highly localized, with a width of $3a$ to $4a$, i.e., it includes only three or four droplets. This width is consistent with the experimental observation of a fluctuation of $1 + 3$ droplets [5, 7].

We note that the spatial width of $\approx 4a$ of this disturbance corresponds to a wave number of $ka \approx \pi/2$. In the wave spectra, this same wave number of $\pi/2$ was found to have a concentration of spectral power. Thus, we can suggest that the concentration of spectral power that is seen in the wave spectra is due to the spatial localization of the wave's growth.

#### 3. Requirement of nonlinearity

We perform a test to assess the role of nonlinearities in the instability's growth. We do this by comparing results for the simulation run with different expressions for the potential: for the fully nonlinear run we use Eq. (3), while for the linear run we use Eq. (4) and retain only the first four terms on the right hand side of Eq. (4).

Results in Fig. 7 show that there is no growth of kinetic energy of the droplets when nonlinearities are turned off. Thus, we can conclude that the instability requires nonlinearities in the potential. This conclusion confirms the suggestion of the experimenters [5, 7] that the instability is essentially the result of a nonlinearity.

#### 4. Requirement of transverse motion

We perform another test to determine whether motion in the longitudinal direction can grow due to an instability in the absence of transverse motion. In this test, we constrain the droplets to move only along the $x$ axis by loading the simulation with non-zero initial displacements only in the $x$ direction. Otherwise, the simulation in this test was the same as for our other fully nonlinear ones.

Results, shown with the open circles in Fig. 7, indicate no growth when the droplets are constrained to move only along the $x$ axis. This finding demonstrates that the instability requires transverse motion.

Considering that all low-amplitude fluctuations can be decomposed as a spectrum of longitudinal waves for motion along the $x$ axis and transverse waves for motion in the $y$ direction, it is reasonable to conclude that the instability involves a coupling between longitudinal and transverse waves. Moreover, combining this conclusion with the previous one for nonlinearity, we have confirmed the suggestion of Beatus *et al.* that *the instability involves an interaction of the longitudinal and transverse waves*. While we have not determined the exact nature of this interaction, one possibility is the three-wave mixing mechanism described in Sec. III B 1.

## IV. SUMMARY

We performed a numerical simulation to study waves and an instability in a 1-D array of water droplets in a microfluidic channel. The droplets were modeled as point objects that interact with each other via a hydrodynamic potential. The oscillatory motion of the droplets, as characterized in the inertial frame of the droplets, exhibits two waves: a longitudinal wave that is backward with respect to the direction of oil flow, and a transverse wave that is forward. The simulation spectra for these waves agree well with the previous experiment that we simulate [5]. The droplet motion increases with time; this instability varies nonexponentially with time, and its growth is spatially localized. We performed two tests that confirm that the instability requires (1) nonlinearities in the hydrodynamic potential and (2) an interaction between longitudinal and transverse waves. A possible candidate for this interaction is a nonlinear three-wave mixing, which is suggested by a feature seen in our wave spectrum.

### Acknowledgments

The authors thank R. Bar-Ziv, T. Beatus and T. Tlusty for helpful discussions. This work was supported by NSF and NASA.

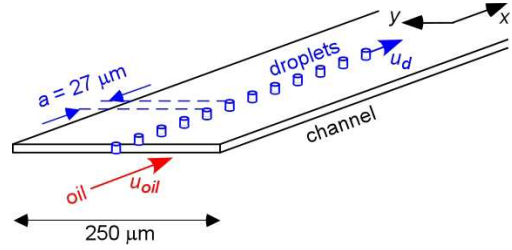

FIG. 1: (Color online) Sketch of a 1-D array of water droplets dragged slowly by a faster flow of oil in a microfluidic channel, as in the experiment of [5].

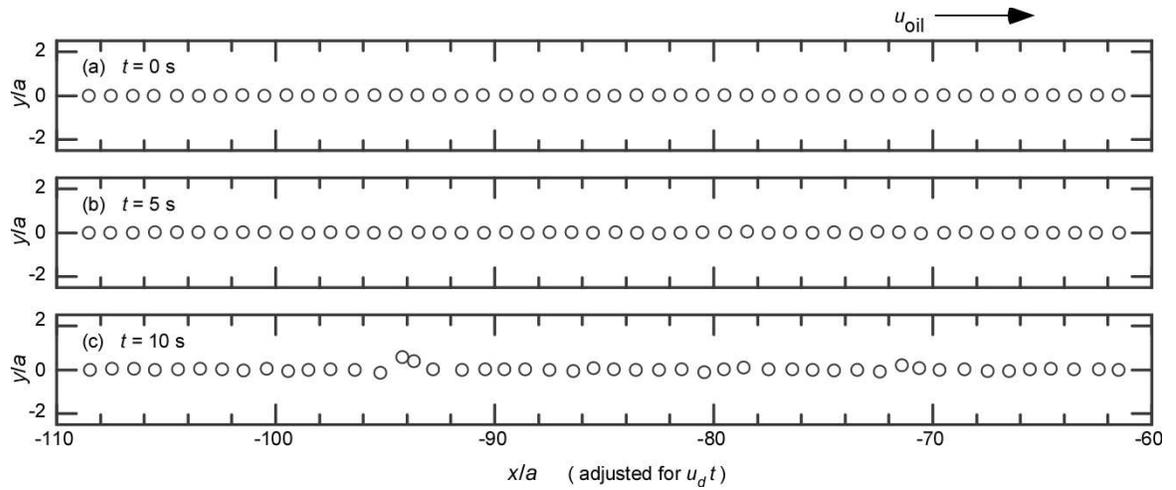

FIG. 2: Droplet positions at three times in the simulation, shown in the inertial frame of the array of droplets. Only a portion of a total of 256 droplets is shown here. (a) At the time $t = 0$, the initial positions of the droplets are specified. We then integrate the fully nonlinear equation of motion, Eqs. (1)-(3), to advance the droplet position and velocity each time step. (b) At $t = 5$ s, a droplet fluctuates about its equilibrium position. (c) At $t = 10$ s, the displacements of the droplets have grown as large as $a$, due to the instability.

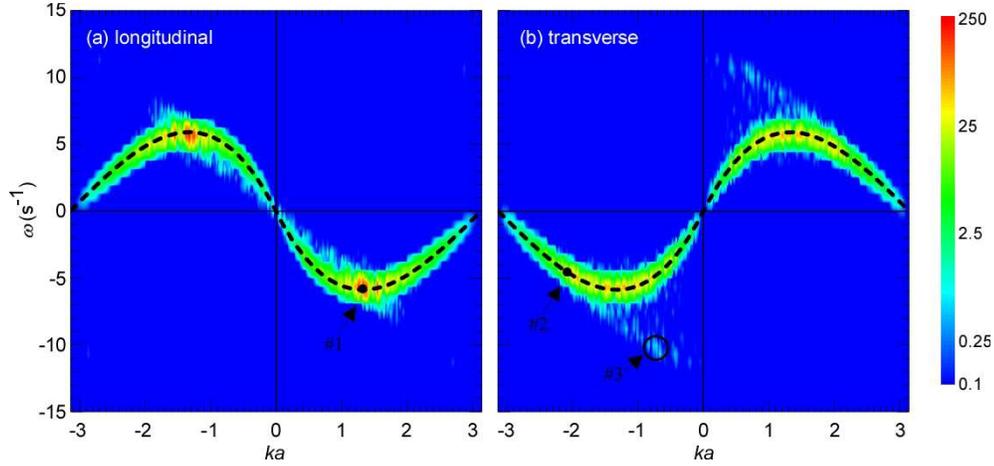

FIG. 3: (Color online) Power spectra of longitudinal (a) and transverse (b) waves, with the logarithm of spectral power shown as a function of frequency and wave number. Spectral power is mostly concentrated in narrow ranges, which are centered on the dispersion relations for a linear theory, Eq. (7), shown as dash lines. The feature near $ka \approx \pm 0.8$ and $\omega \approx \pm 10$ s$^{-1}$, marked #3 in (b), is not predicted by the linear theory, and is interpreted here as an indication of nonlinear effects, possibly three-wave mixing of the waves marked #1 and #2.

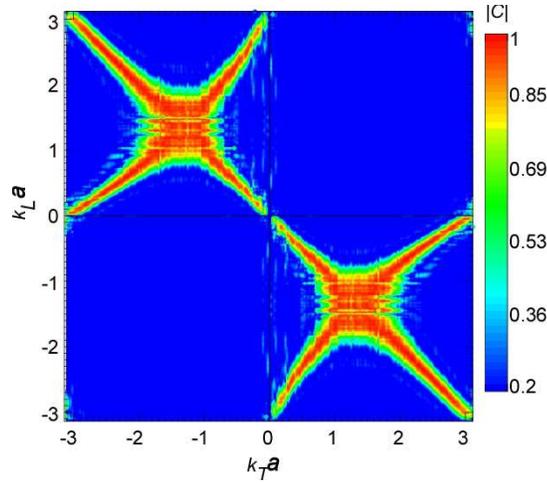

FIG. 4: (Color online) Correlation strength $|C|$, as defined in Eq. (11), for longitudinal $k_L$ and transverse $k_T$ waves. We find a strong correlation, for certain combinations of $k_L$ and $k_T$. The contour scale has a threshold of 0.2 to suppress noise.

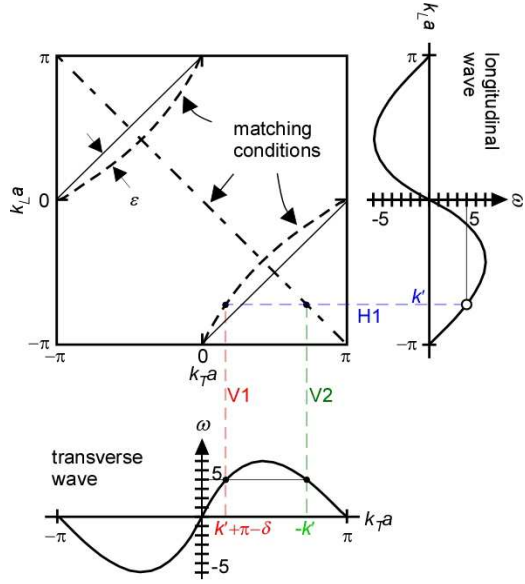

FIG. 5: (Color online) Interpretation of correlation data in Fig. 4. Wave numbers that exhibit high correlations in Fig. 4 obey matching conditions for wave numbers shown with a dot-dash curve with $k_L a = -k_T a$ and dash curves with $k_L a = k_T a \pm (\pi - \epsilon)$ where $\epsilon \ll 1$. We also find a matching condition for frequency, by comparing to the dispersion relations, Eq. (7), which are plotted along the edges of the correlation graph. As an example of these matching conditions, we consider a longitudinal wave $k'$ as marked with an open circle. Following the horizontal line $H1$ across the diagram, strong correlation occurs for the transverse waves with wave numbers $-k'$ and $k' + \pi - \delta$, which have the same frequency as the longitudinal wave.

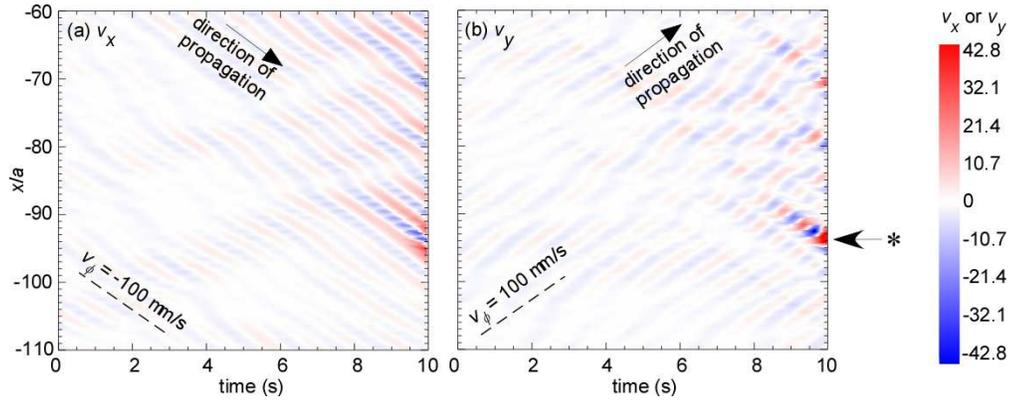

FIG. 6: (Color online) Space-time diagrams for velocities (a) $v_x$ and (b) $v_y$. To make the space-time diagram easier to examine, we present the data for only a portion of the array of droplets, $146 \leq m \leq 196$. The sloped stripes in (a) and (b) indicate that longitudinal and transverse wave packets propagate in opposite directions. Note that the wave growth is spatially localized and that the high-amplitude wave packets at $t > 8$ s include only three or four droplets at the location marked $*$.

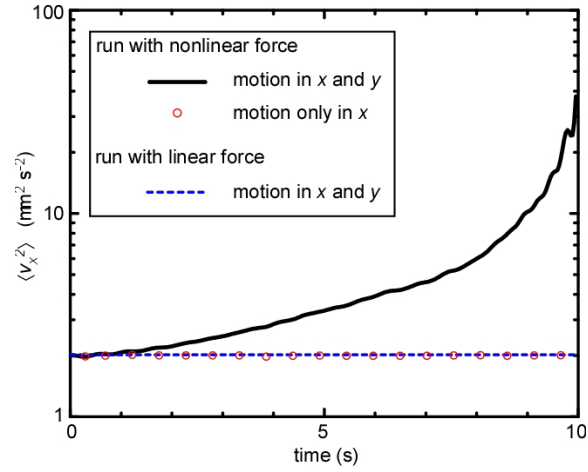

FIG. 7: (Color online) Instability growth, as measured by the time series of the mean squared velocity, averaged over the entire length of the 1-D array. The simulation was run three ways for this test. The solid curve is for the fully nonlinear case using Eq. (3) when integrating the equation of motion, as in Figs. (1)-(5). The dashed curve is for the linear case using Eq. (4). The open circles are results for the fully nonlinear simulation with droplets constrained to move only longitudinally, i.e., in the $x$ direction. We find that for the nonlinear case there is rapid nonexponential growth, but for the other two cases there is no growth, indicating that the instability requires nonlinearity and transverse motion
.